\documentclass[3p]{elsarticle}
\usepackage{hyperref}
\usepackage{amssymb}
\usepackage{amsmath}
\usepackage{graphicx}
\usepackage{multirow,array}
\usepackage{bm}

\begin{document}

\begin{frontmatter}

\title{Fixation probability on clique-based graphs}

\author[JC]{Jeong-Ok Choi}

\author[UY]{Unjong Yu\corref{mycorrespondingauthor}}
\cortext[mycorrespondingauthor]{Corresponding author}
\ead{uyu@gist.ac.kr}

\address[JC]{Division of Liberal Arts and Sciences, Gwangju Institute of Science and Technology, Gwangju 61005, South Korea}
\address[UY]{Department of Physics and Photon Science, Gwangju Institute of Science and Technology, Gwangju 61005, South Korea}

\begin{abstract}
The fixation probability of a mutant in the evolutionary dynamics of Moran process
is calculated by the Monte-Carlo method on a few families of clique-based graphs. 
It is shown that the complete suppression of fixation can be realized
with the generalized clique-wheel graph in the limit of
small wheel-clique ratio and infinite size.
The family of clique-star is an amplifier,
and clique-arms graph changes from amplifier to suppressor
as the fitness of the mutant increases.
We demonstrate that the overall structure of a graph can be more important
to determine the fixation probability
than the degree or the heat heterogeneity.
The dependence of the fixation probability
on the position of the first mutant is discussed.
\end{abstract}


\end{frontmatter}


\section{Introduction}

Evolutionary dynamics studies the spread or extinction of agents
of different feature or trait within a population \cite{n}.
It can be applied to various disciplines such as the opinion
dynamics, the immunology, and the oncology, as well as the ecology \cite{n,Sen13}.
Traditionally, the evolutionary dynamics has been studied in infinite homogeneous populations \cite{m}.
The dynamics starts with one mutant with fitness $r$ as all the rest of the agents are
residents (non-mutants), whose fitness is set to one. 
Each step of the dynamics consists of two sub-steps in the order of birth and death.
An agent to reproduce is chosen randomly proportional to its fitness,
and another agent is chosen uniformly at random to follow the property of the first-chosen agent.
This step is repeated until the mutant or the resident prevails the whole population.
The fixation probability, which is the probability that the mutant
spreads over the whole population, depends on the fitness of the mutant
$r$ and the population size $N$ as Moran's probability $f_M(N,r) = \left( 1-1/r \right)/\left(1-1/r^N \right)$ \cite{m} for homogeneous populations.
We let $f_M(r) = \lim_{N \to \infty} f_M(N,r)$.

Lieberman {\it et al}., however, found that the fixation probability can depend crucially on the 
structure of population, which is represented as a graph \cite{lhn}.
A vertex in a graph represents an agent in the population, and each edge
represents reproductivity between the end vertices of the edge.
Therefore, a homogeneous population is represented by a
complete graph. There are at least two ways to model reproductivity
from one agent to another using graphs:
directed graph ({\it digraph}) vs. (undirected) graph.
In a digraph, each edge has a direction from a vertex $v$ to another vertex $u$,
and it means an agent at $v$ can reproduce into $u$ but not the other way. In a
graph, on the other hand, for every edge both directions are
possible. The set of graphs (with both directions
given on each edge) is a subset of the set of digraphs.
When the fixation probability is larger (smaller) than the Moran's probability
for $r > 1$, the graph is called an amplifier (suppressor).
In the case of a digraph, it is easy to make an efficient amplifier or
suppressor for sufficiently large $N$. It is even possible to make fixation
probability 1 or 0 for $r>1$ using some specific kind of digraphs \cite{lhn,galanis17}.
However, within undirected graphs, it is not trivial to find
such a strong amplifier or a suppressor.
According to the isothermal theorem \cite{lhn,br}, a simple connected graph has
Moran's probability $f_M(N,r)$ if and only if it is regular.
For irregular graphs, it is difficult and complicated to calculate the
fixation probability of a graph analytically, and formula of
the fixation probability is known only for a few graphs. 
The analytic expression of the fixation probability for star graph was derived to be
$f_{\rm star}(N,r) = \left(1-1/r^2\right) / \left(1 - 1/r^{2N} \right) = f_{M}(N,r^2)$
for sufficiently large $N$ \cite{lhn,br}.
The star graph is known to have the strongest amplification
among undirected graphs up to now.
(Recently, in a limited interval for $r$, a family of graphs (comets) with higher fixation probability
than the star graphs is discovered \cite{pavlogiannis17}.)

It is known that the variance of vertex degree (degree heterogeneity) and the fixation probability
are positively correlated \cite{brs2, tl}.
As supporting the result, it is known that most of irregular graphs
are amplifiers \cite{ht} and the amplification
is enhanced by the degree heterogeneity \cite{br}.
Furthermore, in \cite{tl}, the authors proposed the heat heterogeneity $H_t(G)$
to explain the amplification of irregular graphs.
The heat heterogeneity is defined by
\begin{eqnarray}
H_t(G) = \frac{1}{N} \sum_{i = 1}^{N}\left(T_i - \overline{T}\right)^2 ,
\end{eqnarray}
where $T_i$ is temperature of vertex $i$,
which is defined by the sum of inverse degree ($1/d_k$) of its $k$-th neighbor:
$T_i = \sum_{k \in N(i)} (1/d_k)$.
$N(i)$ is the set of neighbors of vertex $i$.
A high-temperature vertex is expected to change its trait more often
than a low-temperature one \cite{n}.
Average temperature $\overline{T}$ is the arithmetic mean of $T_i$.
They tested complete bipartite graphs and randomly sampled general networks,
and showed that heat heterogeneity has better correlation to fixation probability.
These arguments can explain strong amplification of the star graph,
for the star graph has very large degree heterogeneity and heat heterogeneity.
Because all the irregular graphs have larger heterogeneity than the regular graph,
most of them are expected to be amplifiers.
Therefore, it is challenge to find a strong suppressor.

Recently, as opposing the above result,
a promising undirected suppressor was proposed \cite{mnrs}.
It is composed of a clique of size $n$, a ring of the same size,
and $n$ edges connecting the vertices of the clique and those of the ring one-by-one,
which is called as a clique-wheel. [See the middle graph in Fig.~\ref{fig_clique_wheel}(a).]
It was insisted that the fixation probability of the clique-wheel is smaller
than half of the Moran's probability for $1<r<4/3$ as $n \to \infty$.
However, they could not present definite fixation probability.
Also, the result for $r \ge 4/3$ remained open:
It is not clear whether the clique-wheel graph would 
act as a suppressor or an amplifier for $r \ge 4/3$.
The questions that will be answered in this paper are as follows.
(i) Is it possible to suppress the fixation probability completely
    within undirected graphs?
(ii) Is it possible that a graph can be both of the amplifier and suppressor
  depending on the fitness?
(iii) Which properties of a graph are important in the fixation probability?
We show that the fixation probability 
of the original clique-wheel graph
is half of the Moran's probability in $1< r< 4/3$,
and it rapidly approaches the Moran's probability from below for $r \ge 4/3$.
The star and clique-wheel graphs are at each extreme in fixation property.
Interestingly, both can be considered as an inter-connected network composed of core (clique)
and non-core graphs: the star graphs are composed of a clique of size one and many vertices
that are connected to the clique.
In this paper, we generalized these graphs and studied clique-based graphs systematically
by varying the ratio of core/non-core graphs and shape of non-core graphs. 
As a result, we found that it is possible to make the fixation probability zero
in a limited regime, which is the strongest suppressor ever known.
In addition, we found graphs that changes from the amplifier to the suppressor
depending on the mutant fitness.
We also discuss how the fixation probability depends on the starting position of a mutant
in clique-based graphs.

\section{Model and Methods}

The dynamics used in this work is the same as the Moran process, which is explained in the introduction,
except that the graph structure is taken into account. 
The fixation probability depends on the vertex where a mutant is initially
located to start. In this work, the location of the first
mutant was chosen sequentially through all the vertices of the graph.
After the selection of the first mutant, agents to rebirth
and to die are chosen at random just as the standard Moran process.
This method has smaller error than the random selection
of the first mutant, though they give equivalent
results about fixation probability when the simulation is repeated
many times.
The simulation was repeated at least $10^8$ times per each case.
More simulations were performed close to $r=1$, where the fixation probability is small.


In this work, we consider five kinds of graphs that are composed of a clique of size $n$
and non-core graphs connected to the clique.
Let $K_n$, $mK_1$, and $C_m$ be the clique on $n$ vertices, $m$ isolated vertices, and the cycle on $m$ vertices, respectively.
Then the five kinds of families of graphs can be defined as follows.

\begin{figure}
\centerline{\includegraphics[width=12.0cm]{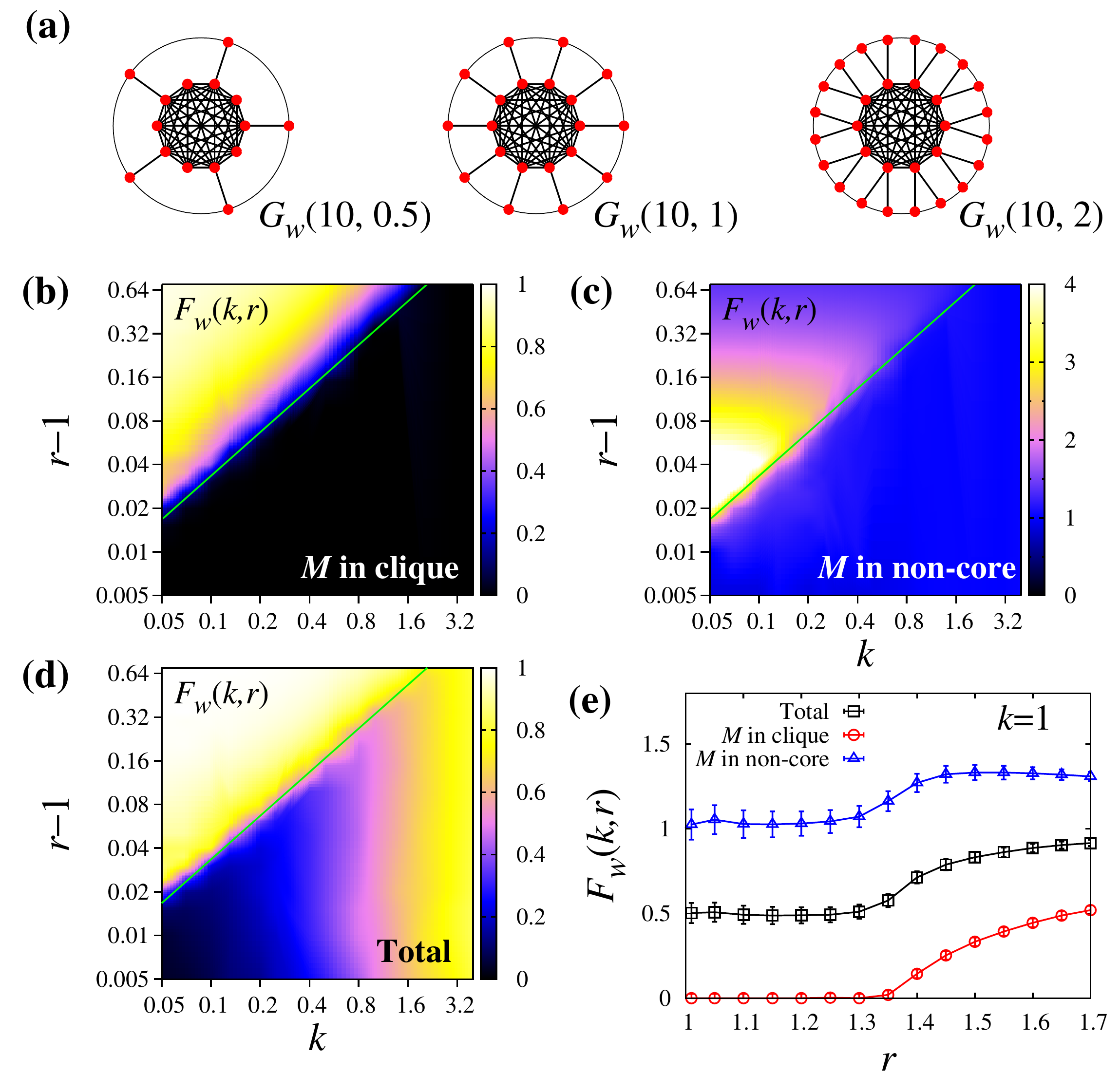}}
\caption{(a) Clique-wheel graphs with different $k$.
         (b-d) Fixation probability of the clique-wheel graphs $f_w(k,r)$ relative to that of the infinite homogeneous graph $f_M(r)=1-1/r$
             when a mutant appears in one of the clique vertices for (b), non-core vertices for (c), and randomly everywhere for (d).
             The green straight line represent $r=1+k/3$.
         (e) Fixation probability divided by $f_M(r)$ as a function of the fitness of the mutant $r$ in the clique-wheel graph with $k=1$
             for a mutant in clique, wheel, and everywhere.
          }
\label{fig_clique_wheel}
\end{figure}

\begin{figure}
\centerline{\includegraphics[width=12.0cm]{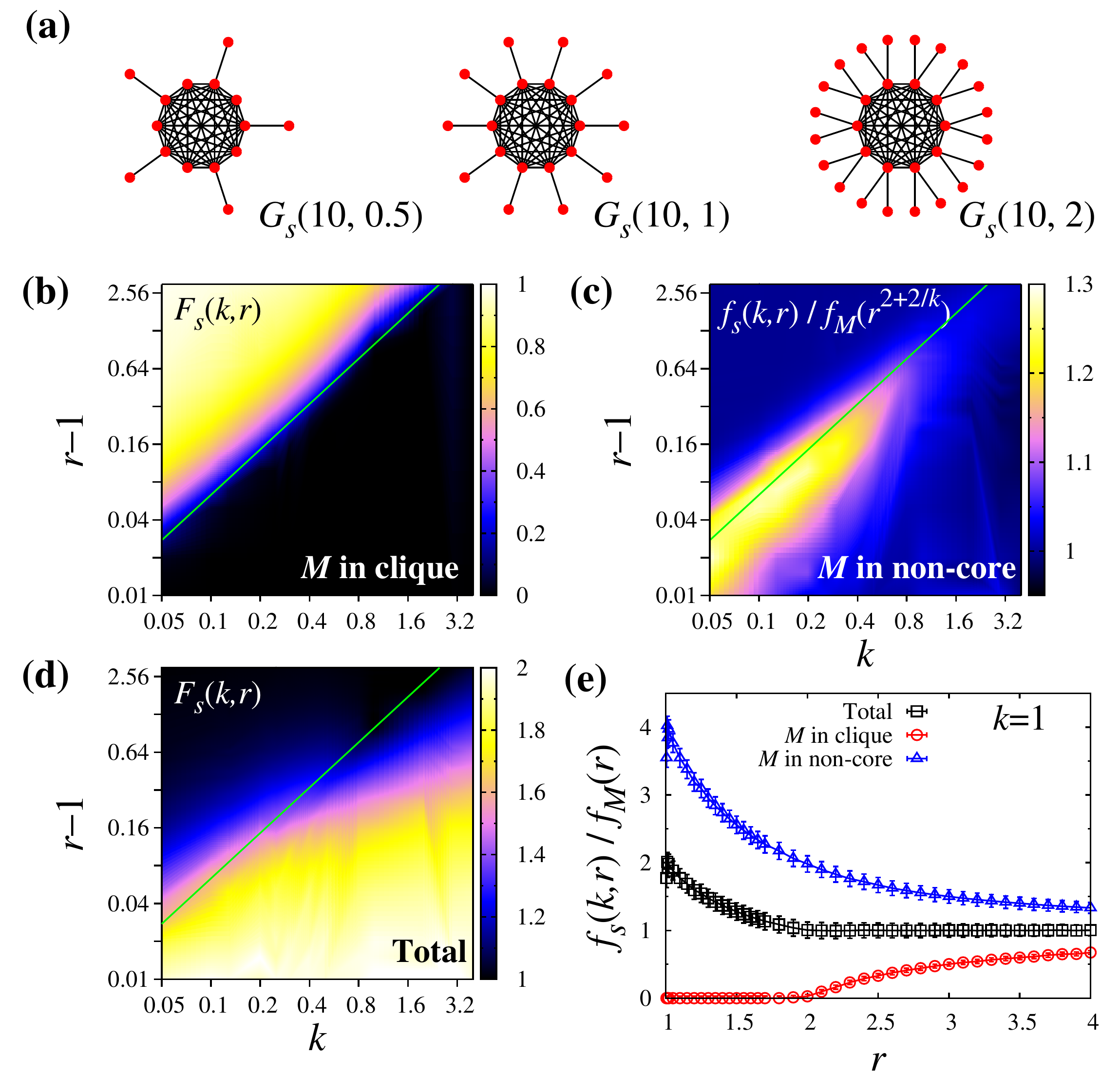}}
\caption{(a) Clique-star graphs with different $k$.
         (b-d) Fixation probability of the clique-star graphs $f_s(k,r)$ relative to that of the infinite homogeneous graph
             $f_M(r^\alpha)=1-1/r^\alpha$ with amplified mutant fitness $r^\alpha$
             when a mutant appears at one of the clique vertices for (b), non-core vertices for (c), and randomly everywhere for (d).
             Note that $\alpha=1$ for (b) and (d), and $\alpha=2+2/k$ for (c).
             The green straight line represent $r=1+k^{1.2}$.
denote
for a mutant in clique, wheel, and everywhere.
          }
\label{fig_clique_star}
\end{figure}

\begin{figure}
\centering{\includegraphics[width=10.0cm]{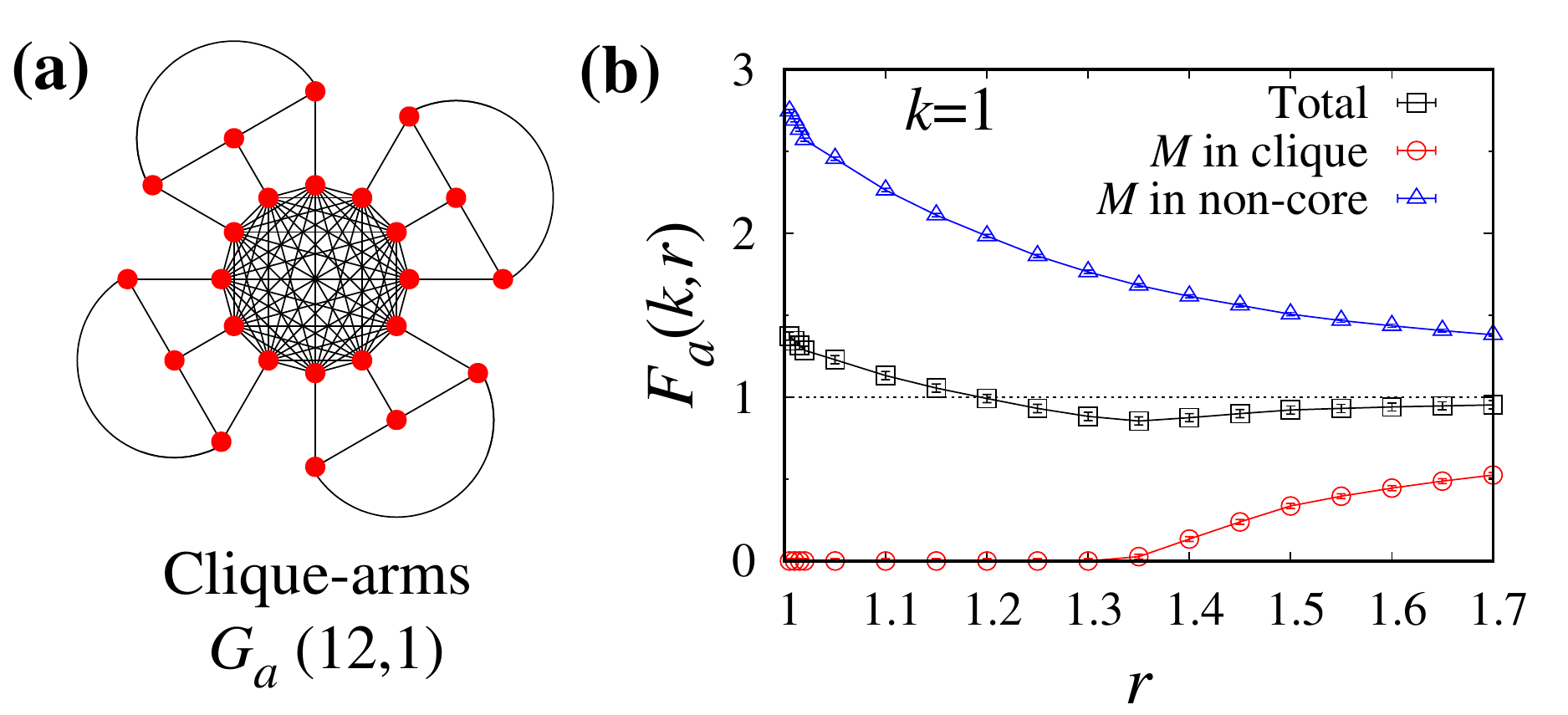}}
\caption{(a) A clique-arms graph with $n=12$ and $k=1$.
         (b) Fixation probabilities divided by $f_M(r)=1-1/r$ as a function of
             the fitness of the mutant $r$ in the clique-arms graph with $k=1$
             for a mutant in clique, wheel, and everywhere.
             The statistical error is smaller than the data point size.
          }
\label{fig_clique_arms}
\end{figure}

\begin{figure}
\centering{\includegraphics[width=12.0cm]{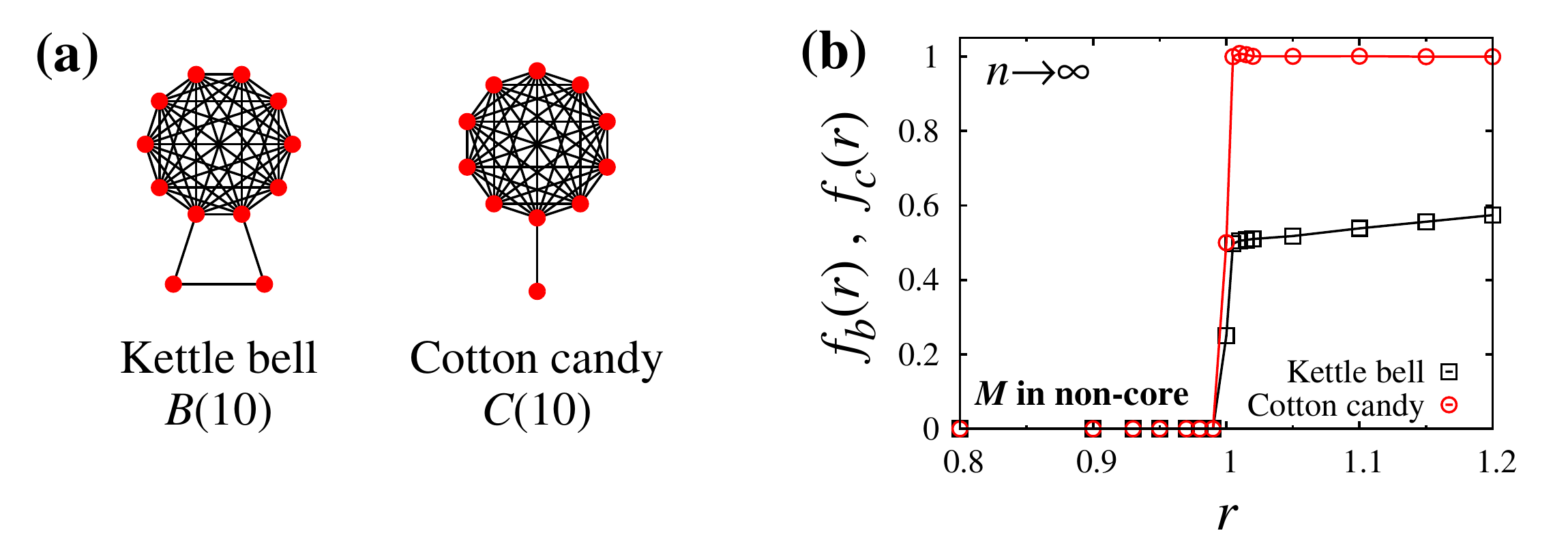}}
\caption{(a) Graphs of a kettle-bell and cotton-candy with $n=10$.
         (b) Fixation probabilities of the kettle-bell and cotton-candy graphs
             when a mutant appears in the non-core vertices
             The statistical error is smaller than the data point size.
          }
\label{fig_cotton_candy}
\end{figure}

\begin{itemize}
\item {\bf Clique-wheel} $G_w(n, k)$: Start with $K_n$ and $C_{nk}$, and insert $nk$ edges evenly between the vertices of
$K_n$ and $C_{nk}$. [See Fig.~\ref{fig_clique_wheel}(a).]
\item {\bf Clique-star} $G_s(n, k)$: Start with $K_n$ and $(nk)K_1$, and insert $nk$ edges evenly between the vertices of
$K_n$ and $(nk)K_1$. [See Fig.~\ref{fig_clique_star}(a).]
\item {\bf Clique-arms} $G_a(n, k)$: From $G_w(n, k)$ replace $C_{nk}$ by 3-cycles or 4-cycles with 3-cycles as many as possible. 
                     [See Fig.~\ref{fig_clique_arms}(a).]
\item {\bf Cotton-candy} $C(n)$: Add one leaf to one vertex of $K_n$. [See Fig.~\ref{fig_cotton_candy}(a).] This graph is a special case of $G_s(n, k)$ with $k=1/n$.
\item {\bf Kettle-bell} $B(n)$: Attach $K_2$ to any two vertices of $K_n$ by adding two independent edges. [See Fig.~\ref{fig_cotton_candy}(a).]
\end{itemize}
As for the clique-arms graph, only $k=1$ case is investigated in this paper.

In all the cases, the fixation probability was calculated for many different sizes 
and the extrapolation to the infinite graph ($n\rightarrow\infty$) was performed.
Typically, seven or more values of the clique size $n$ were chosen from 50 to 1000.
The extrapolation was done by assuming that the fixation probability $f(n,r)$
behaves as $f(n,r)=f(\infty,r)+\lambda_1 (1/n)+\lambda_2 (1/n^2)$ for sufficiently large $n$.
We found that the size dependence of $f(n,r)$ is small, and this tentative formula
fits very well for $n\ge 50$ except very close to $r=1$, where a few small graphs were omitted in the fitting.

\section{Results}

$F(G, r)$ denotes the ratio of the fixation probability over
$f_M(r)$, that is, $f(G, r)/f_M(r)$. 
A graph $G$ is called a {\it suppressor} ({\it amplifier}, resp.) of selection if $F(G, r) < 1$ ($F(G, r) > 1$, resp.)
for any $r > 1$.
We will explain result of the fixation probability in terms of the
effect of the edges connecting the core clique and the rest.
Although the graphs we consider are finite, the results describe
asymptotic behaviors of fixation on graphs with sufficiently large $n$, in
other words, $n \to \infty$ for each family we consider.
A summary of results is listed in Table~\ref{table_all}.
The formulas in the table are estimates for their limits when $k \to \infty$.
For our discussion on main results, $f_{\eta}(n, k, r)$ and $F_{\eta}(n, k, r)$
denote $f(G_{\eta}(n, k), r)$ and $F(G_{\eta}(n, k), r)$, respectively, 
with $\eta = w, s$, and $a$.
We also represent $f(G_{\eta}(\infty, k), r)$ by $f_{\eta}(k, r)$.



\subsection{Family of clique-wheels}
Figure~\ref{fig_clique_wheel} describes the family of general clique-wheel
graphs $G_w(n, k)$ and their fixation probabilities.
All of them are suppressors for $r>1$,
but there are two distinguished regimes.
For $1 < r < 1+k/3$, a mutant that started on the clique is always extinguished
and so the fixation probability is zero.
If a mutant started on the wheel, then the fixation probability is almost the same as Moran
process. Since there are $n$ and $nk$ vertices on the clique and on the wheel, respectively,
$F_w(k, r)$ becomes $k/(k+1)$. For $r > 1 + k/3$, the fixation probability rapidly
increases up to Moran's probability as $r$ increases.
Hence, in particular, $G_w(n, k)$ with very small $k$ shows strong suppression, which is
the strongest known so far. In the limit of $k\to 0$, $r\to 1^{+}$, and $n\to \infty$,
complete suppression ($F_w(n,k,r)=0$) is expected.
For a finite graph, we obtained $F_w(3000,0.05,1.01)=0.094(2)$.
However, the region of strong suppression is very narrow
only close to $r=1$. For larger $k$, the suppression becomes weak, but it is effective in wider regime.

In another limiting case of infinite $k$ with fixed $n$, $G_w(n, k)$ has similar structure
as a wheel graph of infinite size. 
We found that the fixation probability of $G_w(n, k)$ with sufficiently large core (e.g., $n \ge 50$) 
is smaller than the Moran's probability, but it approaches that with $k \to \infty$.
This is consistent with a wheel graph $G_w(1, k)$,
which is an amplifier for finite $k$ but whose fixation probability approaches
the Moran's probability for infinite $k$.

The heat heterogeneity of $G_w(n, k)$ is $[k(n-4+k)^2]/[9(n-1+k)^2]$ for $k \ge 1$.
Although $G_w(n, k)$ is a suppressor, its heat heterogeneity is larger than that of regular graphs, which is zero.
However, there is a positive correlation between the heat heterogeneity and the fixation probability within
this family, for both are increasing function of $k$ with fixed $r$ in $k>3(r-1)$ for sufficiently large $n$.

\subsection{Family of clique-stars}  
Figure~\ref{fig_clique_star} explains the structure of clique-star graphs $G_s(n, k)$ and their fixation probability,
which is always larger than the Moran's probability.
Just like the clique-wheel graph, the fixation probability of a mutant started on the clique is smaller
than the Moran's probability: It is zero in $1 < r < 1 + k^{1.2}$ and increases slowly up to $f_M(r)$ for larger $r$ with infinite $n$.
To the contrary, a leaf has the least degree and is the most advantageous to start for a mutant.
When a mutant starts on a leaf, it becomes amplified with amplification exponent
larger than $2+2/k$. In other words, $f_s(k, r) > f_M(r^{2+2/k})$.
This amplification is so strong to make the total fixation probability
amplified in spite of the suppression by a mutant from the clique.
Since there are $n$ vertices on the clique and $nk$ vertices on leaves,
the total fixation probability should be always 
larger than $[k/(k+1)] f_M(r^{2+2/k})$ and between $f_M(r)$ and $f_{\mathrm{star}}(r)$.
The fixation behavior of $G_s(n, k)$ as $k \to \infty$ converges to the fixation probability of a star graph:
$\lim_{k \to \infty} [k/(k+1)] f_M(r^{2+2/k}) = f_M(r^2)$.

The heat heterogeneity of $G_s(n, k)$ is $\left[{k(n-2+k)^2}\right]/{(n-1+k)^2}$ for integer $k \ge 1$,
which is larger but behaves exactly the same way as the heat heterogeneity of $G_w(n, k)$.
For $k \gg n$, $G_s(n, k)$ is getting close to a star graph and its
heat heterogeneity [$H_t(G_s(n, k)) \sim k$] approaches that of the star graph
with $n+1$ vertices [$H_t(G_s(1, n)) \sim (n-1)^2 / n \sim n$ ].
The fact that both of $F_s(n, k, r)$ and $H_t(G_s(n, k))$ increase as $k$ for $k \gg n$
is consistent with Ref.~\cite{lhn} and explains
why the star graphs are strong amplifiers. 


\subsection{Family of clique-arms}
The family of clique-arms graph $G_a(n, k)$ is interesting because its degree distribution is
exactly the same as the family of the clique-wheels $G_w(n, k)$ for all $n$ and $k$.
Therefore, their degree heterogeneity
and heat heterogeneity are the same as each other.
However, the overall structure of $G_a(n, k)$ can be regarded as a mixture of $G_w(n, k)$ and $G_s(n, k)$.

Figure~\ref{fig_clique_arms} shows the fixation probability of $G_a(n, k)$ with $k=1$.
Like $G_w(n, 1)$, if a mutant starts on the clique then it dies out so the fixation probability is zero for $1 < r < {4}/{3}$.
To the contrary, a mutant that started at any vertex from the non-core fixates with fixation probability more than the Moran's probability.
The amplification is large enough to make overall amplification ($F_a(n, 1, r) > 1$) in $1<r\lesssim 1.2$.
For $r \gtrsim 1.2$, overall fixation probability is suppressed because the fixation probability of a mutant starting at the non-core
is not so strong. Therefore, $G_a(n, k)$ is the first example that has a mixed fixation behavior depending on fitness $r > 1$,
for graphs whose fixation probabilities are known so far has only one kind of fixation behavior: either a suppressor or an amplifier for all $r > 1$.
Although Fig.~\ref{fig_clique_arms} shows the results in the limit of infinite $n$, amplifier-suppressor crossover is still observed for $n=50$.
This result implies that the correlation between the fixation probability and heterogeneity 
may not be a determining factor and the overall structure of a graph can be more important.


\begin{table}
\caption{The fixation probability of the mutant in the clique-wheel, the clique-star, and the clique-arms graphs
depending on the fitness and the location of the first mutant ($M$).
The size of the graphs is infinite ($n\rightarrow\infty$).
S, A, and 0 in the rightmost column represent suppression, amplification,
and neither of them, respectively.}
\label{table_all}
\begin{center}
\begin{tabular}{|c|c|c|c|c|c|}
\hline
Graph & \multicolumn{2}{c|}{Fitness and location of $M$} & Fixation probability & \\
\hline
\multirow{6}{*}{$G_w(\infty,k)$} & \multirow{3}{*}{$1<r<1+k/3$}     & Clique & $f_w(k,r)=0$                       & S \\
                                 &                                  & Non-core & $f_w(k,r)=f_M(r)$                & 0 \\
                                 &                                  &  Total & $f_w(k,r) = [k/(k+1)]f_M(r)$       & S \\
                                 \cline{2-5}
                                 & \multirow{3}{*}{$r>1+k/3$}       & Clique & $f_w(k,r)<f_M(r)$                  & S \\
                                 &                                  & Non-core & $f_w(k,r)>f_M(r)$                & A \\
                                 &                                  &  Total & $[k/(k+1)]f_M(r)<f_w(k,r)<f_M(r)$  & S \\
\hline
\multirow{6}{*}{$G_s(\infty,k)$} & \multirow{3}{*}{$1<r<1+k^{1.2}$} & Clique & $f_s(k,r)=0$                       & S \\
                                 &                                  & Non-core & $f_s(k,r)>f_M(r^{2+2/k})$        & A \\
                                 &                                  & Total  & $f_s(k,r)>\max\{f_M(r),[k/(k+1)]f_M(r^{2+2/k})\}$ & A \\
                                 \cline{2-5}
                                 & \multirow{3}{*}{$r>1+k^{1.2}$}   & Clique & $f_s(k,r)<f_M(r)$                  & S \\
                                 &                                  & Non-core & $f_s(k,r)>f_M(r^{2+2/k})$        & A \\
                                 &                                  & Total  & $f_s(k,r)>\max\{f_M(r),[k/(k+1)]f_M(r^{2+2/k})\}$ & A \\
\hline
\multirow{3}{*}{$G_a(\infty,1)$} & \multirow{3}{*}{$r>1$}           & Clique & $f_a(1,r)<f_M(r)$                  & S \\
                                 &                                  & Non-core & $f_a(1,r)>f_M(r)$                & A \\
                                 &                                  & Total  &                                    & S or A \\
\hline
\end{tabular}
\end{center}
\end{table}

\subsection{Cotton candy and kettle bell graphs} 
Prior to the introduction to degree heterogeneity, in fact, Broom et. al. \cite{brs} tested all the graphs
with the number of vertices up to 8 for the effect of location of a starting mutant on the fixation probability.
They confirmed precedent statistical results \cite{brs2,ars} that starting at a smaller-degree vertex is
more advantageous for a mutant to fixate. The results on clique-based graphs confirm the argument.
As shown in Table~\ref{table_all}, the fixation probability is always bigger when a mutant starts
at non-core vertices, which has smaller degree than vertices on the clique.
Furthermore the amplification of non-core vertices becomes stronger as $k$ decreases.

In the cotton-candy graph, which is the limiting cases of $k\to 0$ for the clique-star,
the fixation probability of a mutant from a stick (non-core) vertex
becomes the Heaviside step function $\theta(r-1)$ in the limit of $n\to \infty$,
as shown in Fig.~\ref{fig_cotton_candy}(b).
Some methods for analytic solutions have been developed using Markov chains and Martingales \cite{mgp},
but obtaining the exact solutions is very complicated and only a few fixation probabilities are known. 
However, this behavior can be explained as follows.
When there is only one mutant on the leaf (non-core) vertex, there are two possible
next states: one more mutant in the clique and extinction.
When the mutant is selected first, the first case is realized.
For the second case to be realized, the resident that is connected to the mutant
should be selected first and the mutant should be selected to death.
Therefore, probabilities of the two cases are $rn/(rn+1)$ and $1/(rn+1)$, respectively.
In the limit of $n\to \infty$, the second probability becomes zero.
In other words, a mutant at a leaf will almost always affect the vertex on the clique,
but it is rarely extinguished. After one mutant appear in the clique,
the fixation probability can be described approximately by the Moran's probability.
It is finite for mutant fitness $r$ larger than 1, and zero for $r\le 1$ with $n\to \infty$.
As long as it is not zero, the fixation will be achieved eventually, because
the mutant at a leaf can make its neighbor mutant again and again.



As for the kettle-bell graph, a similar jump exists at $r=1$ when a mutant starts in the handle (non-core),
but $f_b(r=1)=0.25$ and $f_b(r=1^+)=0.5$. $f_b(r)$ increases above 0.5 for $r>1$.
This is because there are two vertices in the handle, which dominates the fixation.
For $r > 1$, the influence on the core by the mutant on the handle is greater
than that by the resident on the handle. Moreover, as $r$ increases this influence by the mutant also increases.
Thus we can see that the roles of non-core of kettle bell and cotton candy graphs are almost the same:
only a mutant at the non-core can fixate.


\section{Summary}
In this paper, we simulated fixation behaviors over the various families of graphs composed of two parts: clique (core) and non-clique (non-core).
Three types for non-core are considered. They are ({\it i}) isolated vertices (clique-stars), ({\it ii}) a big cycle (clique-wheels),
and ({\it iii}) 3-cycles and 4-cycles (clique-arms).
The first two cases are generalizations of known suppressor (clique-wheel with $k=1$) and amplifier (star).
Fixation probabilities were calculated numerically for various ratios of the sizes of core and non-core for the families of graph.
We found that a clique-wheel at very small $k$ is the most strong suppressor as known.
We also discovered the first graph that changes from an amplifier to a suppressor with the mutant fitness $r$ in $r>1$.
The effect of a starting location of a mutant on the fixation probability was also investigated.
For each family from our construction, a mutant starting at the non-core fixates more than a mutant starts at the clique.
Analytic explanation about this fixation behavior is presented.
Finally, a correlation between heat heterogeneity and the fixation probability is discussed.
Contrary to an existing belief, we constructed a counter-example, which shows that there are infinitely many pairs of graphs
with the same heterogeneity such that one graph is a suppressor and the other is an amplifier (clique-wheel and clique-arms).
We propose that an overall structure of graphs can be more important in the fixation probability than the heterogeneity.

\section*{Acknowledgments}
This work was supported by the National Research Foundation of Korea, Grant \# NRF2013R1A1A1076120.
This work was supported by GIST Research Institute (GRI) grant funded by the GIST in 2017.

\section*{References}

\bibliography{ed_bib}

\begin{thebibliography}{10}
\expandafter\ifx\csname url\endcsname\relax
  \def\url#1{\texttt{#1}}\fi
\expandafter\ifx\csname urlprefix\endcsname\relax\def\urlprefix{URL }\fi
\expandafter\ifx\csname href\endcsname\relax
  \def\href#1#2{#2} \def\path#1{#1}\fi

\bibitem{n}
M.~A. Nowak, Evolutionary dynamics, Harvard University Press, Cambridge, MA,
  2006.

\bibitem{Sen13}
P.~Sen, B.~K. Chakrabarti, Sociophysics: An Introduction, Oxford University
  Press, New York, 2013.

\bibitem{m}
P.~A.~P. Moran, Random processes in genetics, in: Proceedings of the Cambridge
  Philosophical Society, Vol.~54, 1958, p.~60.
\newblock \href {http://dx.doi.org/10.1017/S0305004100033193}
  {\path{doi:10.1017/S0305004100033193}}.

\bibitem{lhn}
E.~Lieberman, C.~Hauert, M.~A. Nowak, Evolutionary dynamics on graphs, Nature
  433~(7023) (2005) 312--316.
\newblock \href {http://dx.doi.org/10.1038/nature03204}
  {\path{doi:10.1038/nature03204}}.

\bibitem{galanis17}
A.~Galanis, A.~G\"{o}bel, L.~A. Goldberg, J.~Lapinskas, D.~Richerby, Amplifiers
  for the {M}oran process, J. ACM 64~(1) (2017) 5:1--5:90.
\newblock \href {http://dx.doi.org/10.1145/3019609}
  {\path{doi:10.1145/3019609}}.

\bibitem{br}
M.~Broom, J.~Rycht{\'a}{\v{r}}, An analysis of the fixation probability of a
  mutant on special classes of non-directed graphs, in: Proc. R. Soc. A, Vol.
  464, The Royal Society, 2008, pp. 2609--2627.
\newblock \href {http://dx.doi.org/10.1098/rspa.2008.0058}
  {\path{doi:10.1098/rspa.2008.0058}}.

\bibitem{pavlogiannis17}
A.~Pavlogiannis, J.~Tkadlec, K.~Chatterjee, M.~A. Nowak, Amplification on
  undirected population structures: Comets beat stars, Sci. Rep. 7 (2017) 82.
\newblock \href {http://dx.doi.org/10.1038/s41598-017-00107-w}
  {\path{doi:10.1038/s41598-017-00107-w}}.

\bibitem{brs2}
M.~Broom, J.~Rycht{\'a}{\v{r}}, B.~Stadler, Evolutionary dynamics on graphs-the
  effect of graph structure and initial placement on mutant spread, J. Stat.
  Theory. Pract. 5~(3) (2011) 369--381.
\newblock \href {http://dx.doi.org/10.1080/15598608.2011.10412035}
  {\path{doi:10.1080/15598608.2011.10412035}}.

\bibitem{tl}
S.~Tan, J.~L{\"u}, Characterizing the effect of population heterogeneity on
  evolutionary dynamics on complex networks, Sci. Rep. 4 (2014) 5034.
\newblock \href {http://dx.doi.org/doi:10.1038/srep05034}
  {\path{doi:doi:10.1038/srep05034}}.

\bibitem{ht}
L.~Hindersin, A.~Traulsen, Most undirected random graphs are amplifiers of
  selection for birth-death dynamics, but suppressors of selection for
  death-birth dynamics, PLoS Comput. Biol. 11 (2015) e1004437.
\newblock \href {http://dx.doi.org/10.1371/journal.pcbi.1004437}
  {\path{doi:10.1371/journal.pcbi.1004437}}.

\bibitem{mnrs}
G.~B. Mertzios, S.~Nikoletseas, C.~Raptopoulos, P.~G. Spirakis, Natural models
  for evolution on networks, Theor. Comput. Sci. 477 (2013) 76--95.
\newblock \href {http://dx.doi.org/10.1016/j.tcs.2012.11.032}
  {\path{doi:10.1016/j.tcs.2012.11.032}}.

\bibitem{brs}
M.~Broom, J.~Rycht{\'a}{\v{r}}, B.~Stadler, Evolutionary dynamics on
  small-order graphs, J. Interdiscip. Math. 12~(2) (2009) 129--140.
\newblock \href {http://dx.doi.org/10.1080/09720502.2009.10700618}
  {\path{doi:10.1080/09720502.2009.10700618}}.

\bibitem{ars}
T.~Antal, S.~Redner, V.~Sood, Evolutionary dynamics on degree-heterogeneous
  graphs, Phys. Rev. Lett. 96~(18) (2006) 188104.
\newblock \href {http://dx.doi.org/10.1103/PhysRevLett.96.188104}
  {\path{doi:10.1103/PhysRevLett.96.188104}}.

\bibitem{mgp}
T.~Monk, P.~Green, M.~Paulin, Martingales and fixation probabilities of
  evolutionary graphs, Proc. R. Soc. A 470~(2165) (2014) 20130730.
\newblock \href {http://dx.doi.org/10.1098/rspa.2013.0730}
  {\path{doi:10.1098/rspa.2013.0730}}.

\end{thebibliography}

\end{document}